\documentclass[floatfix,pra,twocolumn,showpacs,amsmath,amssymb,letterpaper,groupaddresses,superscriptaddress]{revtex4}
\usepackage{times}
\usepackage{latexsym}
\usepackage{graphicx}
\usepackage{amsmath,amssymb}
\usepackage{verbatim,times,bbm}

\begin{document}

\title{Quantum reading capacity under thermal and correlated noise}

\author{Cosmo Lupo}
\affiliation{School of Science and Technology, University of Camerino, Via Madonna delle Carceri 9, I-62032 Camerino, Italy}
\affiliation{School of Science and Technology, University of Camerino, Via Madonna delle Carceri 9, I-62032 Camerino, Italy}

\author{Stefano Pirandola}
\affiliation{Department of Computer Science, University of York, Deramore Lane, York YO10 5GH, UK}

\author{Vittorio Giovannetti}
\affiliation{NEST, Scuola Normale Superiore and Istituto Nanoscienze-CNR, Piazza dei Cavalieri 7, I-56126 Pisa, Italy}

\author{Stefano Mancini}
\affiliation{School of Science and Technology, University of Camerino, Via Madonna delle Carceri 9, I-62032 Camerino, Italy}
\affiliation{INFN-Sezione di Perugia, Via A. Pascoli, I-06123 Perugia, Italy}

\begin{abstract}
Quantum communication theory sets the maximum rates at which
information can be encoded and decoded reliably given the physical
properties of the information carriers. 
Here we consider the problem of readout of a digital optical memory, 
where information is stored by means of the optical properties of the 
memory cells that are in turn probed by shining a laser beam on them.
Interesting features arise in the regime in which the probing
light has to be treated quantum mechanically. 
The maximum rate of reliable readout defines the {\it quantum reading capacity}, which
is proven to overcome the classical reading capacity --- obtained by probing with classical 
light --- in several relevant settings.
We consider a model of optical memory in which information is encoded in the (complex-valued)
attenuation factor and study the effects on the reading rates of thermal and correlated noise.
The latter type of noise arises when the effects
of wave diffraction on the probing light beam are taken into account. 
We discuss the advantages of quantum reading over the classical one and show that the former 
is substantially more robust than the latter under thermal noise in the regime of low power per pulse.
\end{abstract}


\maketitle

\section{Introduction}\label{intro}

Data storage devices make use of different stable or metastable states of
suitable physical systems, typically organized in arrays of memory cells, which are
employed to store classical information. In turn, the process of retrieving the stored information
involves a second physical system used to probe the state of the cells.
The readout process is hence made of three steps: 1) the probing system is prepared in
a given initial state; 2) the memory cells and the probe interact according to the
their physical properties; 3) the probe is collected and measured
in order to extract the information encoded in the state of memory cells mirrored
in the final state of the probing system.
The prototypical example is that of digital optical memories, such
as CD's, DVD's and BD's, where information is stored in digital
form according to the optical properties of the memory cells. In
this case, the readout process consists of shining a laser on the
memory cells, whose reflected beam is then collected and properly
measured. Under standard conditions, the use of classical coherent
light with a macroscopic number of photons is sufficient to
guarantee a faithful readout process with negligible probability
of error. On the other hand, this is no longer true in the regime
of weak signals in which quantum fluctuations become predominant
and, due to the uncertainty principle, forbid the faithful
discrimination of coherent states. Remarkably, as it was first
pointed out in~\cite{Pirs}, it is in the few photon regime that
{\it nonclassical} states of light can provide the optimal
probing states for the readout of a classical memory, hence
overcoming the performances of classical light sources. 
This result, which is already interesting by its own, may open the 
way to novel applications of quantum optics for data storage technologies,
e.g., by increasing the data storage capacity or the readout rate
or by allowing the safe and faithful readout of photodegradable
memories.

A model for studying the readout of digital classical memories in
the quantum regime was first considered in~\cite{Pirs}. There, a
model of memory was introduced in which information is encoded in
a binary fashion according to the optical reflectivity of each
memory cell. That is, a low or high reflectivity of the memory
cell is used to store one bit of classical information. Hence,
depending on the reflectivity value, a beam of light impinging on
a memory cell results in an attenuated amplitude of the reflected
beam. It follows that the problem of the memory readout is
naturally formulated as the problem of {\it statistical
discrimination} between two {\it lossy bosonic
channels}~\cite{Pirs,Weedbrook}, and its probability of success
arises as a suitable performance quantifier. Variants of this
model include the cases in which the interaction with the memory
cell affects other optical properties of the reflected beam, as
phase~\cite{Hirota} or both phase and amplitude~\cite{Nair}.
Other variants consider the case where both the ports of the
{\it beam-splitter} memory cell are available to the probing light, with 
the problem of readout becoming equivalent to a discrimination of unitaries~\cite{DAriano1,DAriano2},
see also~\cite{DallArnoPhD,DallArno}.

Moving beyond the case of the readout of a single bit of
information encoded in a memory cell, one has to consider the more
realistic setting for data storage devices in which information is
encoded in (long) arrays of memory cells. 
Classical error correcting codes are also used to further reduce the probability
of faulty readout. General coding theorems (see, e.g.,~\cite{W11})
can be applied to this setting to provide expressions for the
maximum rates at which information can be {\it reliably} retrieved
in the reading process. In particular, when the memory cells are
probed by a quantum system, the asymptotically optimal and
reliable readout rate is given in terms of the {\it Holevo
information}~\cite{Holevo}. In this context one can distinguish two
main settings defined by the physical properties of the probing
states. In the first one the quantum probe is initialized in a
`classical' state, that is, a state characterized by a positive
Sudarshan-Glauber quasidistribution~\cite{Sudarshan}. Classical
states can be written as convex combinations of coherent states
and have a well defined classical limit for $\hbar \to 0$. On the
other hand, nonclassical states are those that show typical
quantum features, for instance squeezing, entanglement and non-locality. 
In both cases, we allow the presence of an ancillary system in order to exploit
(classical or quantum) correlations with the probe. The {\it
classical reading capacity} is hence defined as the maximum rate
of reliable memory readout achievable when the state of the probe
and ancillary systems is a classical one. 
More generally, the {\it quantum reading capacity} is the maximum rate over 
arbitrary quantum states, including nonclassical ones~\cite{QRC}. An
explicit expression for the classical reading capacity was
computed in~\cite{QRC}, as a function of the mean photon number of
the probing light. While the calculation of the quantum reading
capacity remains an open problem, several examples were provided
of nonclassical states outperforming all the classical ones~\cite{QRC},
hence proving that the quantum reading capacity is strictly
greater than its classical counterpart in several settings.

Here we examine the quantum reading capacity of a classical
digital memory in the presence of different kinds of noise. 
For given mean photon number impinging on
each memory cell, we compare the classical reading capacity with that
achievable by an entangled state of the probe and ancillary
systems, which possess Einstein-Podolsky-Rosen (EPR)
correlations. First, we provide analytical expressions in the
presence of thermal background in the limit of faint signals,
showing that the reading rate with the entangled EPR state probing
is unaffected by thermal-like noise. Second, we provide bounds on
the quantum reading rates in the presence of correlated noise
due to light diffraction~\cite{diffraction}. Indeed, wave diffraction of the
reflected light causes cross talks among signals scattered by
neighboring memory cells. Notwithstanding a potential reduction of
the reading rates, we show that the separation between classical
and quantum probing persists in the presence of this kind of
correlated noise.

The article proceeds as follows. 
In Sec.~\ref{sec:OQRC} we review the quantum reading in the optical framework and in
Sec.~\ref{sec:QThermal} we analyze the case of quantum reading in
the presence of background thermal noise. 
In Sec.~\ref{sec:interbit} we introduce and characterize the effects
of diffraction in quantum reading. 
Finally, Sec.~\ref{conclusions} contains conclusions and
remarks. Appendix~\ref{diffract} contains detailed calculations
regarding the quantum description of diffraction.

\section{Quantum reading capacity of optical memories}\label{sec:OQRC}

We consider a model of classical optical memory consisting of a {\it long} array of $K \gg 1$ memory cells,
where the $j$th cell encodes the $d$-ary variable $u(j)$ according to its optical
reflectivity. Hence, a probing beam of light impinging on the $j$th cell will be attenuated by
a complex-valued factor $z_{u(j)}$, describing both attenuation and delay.
In order to optimize the memory performance we allow information to be encoded by means of
a set of $2^C$ distinct codewords $\mathbf{u}^i = (u^i(1), u^i(2), \dots, u^i(K))$ of length $K$, with
$i=1, \dots, 2^C$, at a rate of $C/K$ bits per cell.
These codewords define a classical error correcting code for the memory.
A quantum description of the interaction between the $j$th memory cell and the probing
light is provided by an associated quantum channel $\phi_{u(j)}$ which maps the
incoming beam of light into an attenuated one and accounts for the
presence of background (thermal) noise.
A single mode of incoming light, described by the canonical ladder operators
$a$, $a^\dag$, is hence transformed into an outgoing one according to the
Heisenberg-picture transformation,
\begin{equation}\label{channel}
a \to z_{u(j)} \, a + \sqrt{1-|z_{u(j)}|^2} \, v + \nu \, ,
\end{equation}
where $v$ denotes an environmental `vacuum' mode, and $\nu$ is a Gaussian distributed
classical random variable, with zero mean and variance $n_\mathrm{th}$, modeling the background
thermal noise (note that an alternative way to introduce thermal noise is to
consider an environmental thermal mode instead of a vacuum mode~\cite{Pirs}).
Associated with each codeword of the error correction code there is a
sequence of single-mode quantum channels $\phi_{u^i(1)} \otimes \phi_{u^i(2)}
\otimes \cdots \phi_{u^i(K)}$. We define the marginal ensemble of
quantum channels $\Phi = \{ p_u, \phi_u \}$, where $p_u$ is the
relative fraction of instances of the channel $\phi_{u}$ among the
codewords. The ensemble $\Phi$ is also called the marginal cell of
the memory~\cite{QRC}.

Given the memory model, a reading protocol is characterized by the physical properties
of the probing light.
Here we consider a setting in which each memory cell is independently probed by a collection of
$s$ bosonic modes~\cite{NOTE}, described by the ladder operators $\{ a_k, a_k^\dag \}_{k=1,\dots,s}$,
which are jointly measured with $r$ ancillary modes, which are in turn associated
with the operators $\{ b_{k'}, b_{k'}^\dag \}_{k'=1,\dots,r}$.
We refer to the collective state of the $s+r$ modes as the {\it transmitter}, and denote it as $\rho(s,r)$.
The state of the transmitter after the interaction with a sub-array of cells encoding the codeword
$\mathbf{u}^i$ reads
\begin{equation}\label{ostate}
\rho_{\mathbf{u}^i}(s,r) = \bigotimes_{j=1}^K \rho_{u^i(j)}(s,r)= \bigotimes_{j=1}^K (\phi_{u^i(j)}^{\otimes s} \otimes I^{\otimes r})[\rho(s,r)] \, ,
\end{equation}
where $I$ is the identity transformation acting on each of the ancillary modes.
Finally, the last step of the reading protocol consists of
performing a collective measurement on the outgoing light in order
to discriminate among different codewords. The discrimination can
be performed flawless in the limit $K \to \infty$ by choosing
optimal codewords and collective measurements up to a rate given
by the Holevo information~\cite{Holevo}
\begin{equation}\label{HI}
\chi[\Phi|\rho(s,r)] = S \left[ \sum_u p_u \rho_u(s,r) \right] - \sum_u p_u S \left[ \rho_u(s,r) \right] \, ,
\end{equation}
where $S(\cdot) = - \mathrm{Tr}[(\cdot)\log_2(\cdot)]$ denotes the von Neumann entropy
and $\rho_u(s,r) = (\phi_{u}^{\otimes s} \otimes I^{\otimes r})[\rho(s,r)]$.

The quantum reading capacity of a classical memory with marginal
cell $\Phi$ is finally defined by optimizing the reading rate
$\chi[\Phi|\rho(s,r)]$ over the choice of the probing state
$\rho(s,r)$~\cite{QRC}:
\begin{equation}
C(\Phi) = \sup_{s,r} \sup_{\rho(s,r)} \chi[\Phi|\rho(s,r)] \, .
\end{equation}
However, one has to notice that $\chi[\Phi|\rho(s,r)]$ is upper
bounded by the Shannon entropy of the ensemble $\Phi$, $H(\Phi) =
-\sum_u p_u \log_2{p_u}$, which in turn can be made equal to $\log_2{d}$
by letting $p_u$ to be the flat distribution. It is easy to show
that this bound is saturated by choosing $\rho(s,r)$ to be a pure
state and allowing $s$ to be arbitrarily large~\cite{QRC}. 
For this reason, the quantum reading capacity is a a nontrivial notion
only when we optimize the transmitter state under a suitable
constraint. In the framework of optical readout, the most meaningful physical
constraint is given by fixing the mean number of photons impinging on
each memory cell. 
Thus, we consider fixed-energy transmitters $\rho(s,r,n)$ defined as those
transmitters $\rho(s,r)$ which irradiate an average of $n$ photons on each
memory cell, i.e., such that
\begin{equation}
\mathrm{Tr} \left[ \rho(s,r,n) \sum_{k=1}^s a_k^\dag a_k \right] = n \, .
\end{equation}
We hence define the {\it constrained quantum reading capacity} by optimizing
the reading rate over all the transmitters at fixed energy~\cite{QRC},
\begin{equation}
C(\Phi|n) = \sup_{s,r} \sup_{\rho(s,r,n)} \chi[\Phi|\rho(s,r,n)] \, .
\end{equation}
Note that an alternative (local) energy
constraint consists of fixing the mean number of photons per
signal mode, as recently adopted in Refs.~\cite{local,Tej} --- see~\cite{Weedbrook}
for more details on global and local energy constraints in quantum channel
discrimination.
Also note that a different definition of optical reading capacity has been
considered in Ref.~\cite{SaikatLAST}, where an optimization on the marginal cell is implicitly considered.

In the following we evaluate bounds on $C(\Phi|n)$ by computing
$\chi[\Phi|\rho(s,r,n)]$ for different choices of the transmitter. 
First, we consider the case of `classical' states --- those
having a positive Sudarshan-Glauber
quasidistribution~\cite{Sudarshan} --- of $s+r$ modes. By
restricting to classical transmitters, denoted as
$\rho_\mathrm{c}(s,r,n)$, one defines the constrained {\it
classical reading capacity}:
\begin{equation}
C_\mathrm{c}(\Phi|n) = \sup_{s,r} \sup_{\rho_c(s,r,n)} \chi[\Phi|\rho_c(s,r,n)] \, .
\end{equation}
Classical states form a convex set whose extremal points are
coherent states of $s+r$ modes. It was proven in~\cite{QRC} that
for the noiseless case ($n_\mathrm{th}=0$) the optimal classical
probing state is a single-mode coherent state with $s=1$ and
$r=0$. Here we conjecture, supported by numerical evidence, that
the same holds true even in the noisy case ($n_\mathrm{th} > 0$).
Second, we evaluate $\chi[\Phi|\rho(s,r,n)]$ for an exemplary
`non-classical' transmitter given by 
\begin{equation}
\rho_\mathrm{EPR}(s,s,n) = (|\xi\rangle\langle\xi|)^{\otimes s} \, ,
\end{equation}
where 
\begin{equation}
|\xi\rangle = (\cosh{\xi})^{-1} \sum_{m=0}^\infty (\tanh{\xi})^m |m\rangle_{S_k} |m\rangle_{R_k}
\end{equation}
with $\xi=\mathrm{arc}\sinh{\sqrt{(n/s)}}$ and
$|m\rangle_{S_k} = (m!)^{-1/2}(a_k^\dag)^m |0\rangle$,
$|m\rangle_{R_k} = (m!)^{-1/2}(b_k^\dag)^m |0\rangle$ ($|0\rangle$
denoting the vacuum state). The transmitter
$\rho_\mathrm{EPR}(s,s,n)$ is the tensor product of $s$ EPR
states, providing the simplest description of the output of
parametric down conversion, see e.g.~\cite{Konrad}, as well as the
prototypical example of entangled state in the continuous-variable
setting~\cite{Weedbrook,Paris}.

\section{Quantum reading capacity under thermal noise}\label{sec:QThermal}

We start by considering the case of coherent state transmitters and introduce the quantity
\begin{equation}
C_\mathrm{coh}(\Phi|n) = \sup_{s,r} \sup_{\rho_\mathrm{coh}(s,r,n)} \chi[\Phi|\rho_\mathrm{coh}(s,r,n)] \, ,
\end{equation}
where
\begin{equation}
\rho_\mathrm{coh}(s,r,n) = \bigotimes_{k=1}^s |\alpha_k\rangle\langle\alpha_k| \bigotimes_{{k'}=1}^r |\beta_{k'}\rangle\langle\beta_{k'}| \, ,
\end{equation}
is a coherent state of $s+r$ modes and the photon-number constraint reads $\sum_{k=1}^s |\alpha_k|^2 = n$.

First of all, we notice that since the state $\rho_{\mathrm{coh}}(s,r,n)$ is in the form of a 
direct product between the state of the probe and the state of the ancillary 
system, the subadditivity of the Holevo information implies that the presence of the ancillary
modes cannot increase the reading rate, that is,
\begin{equation}
C_\mathrm{coh}(\Phi|n) = \sup_{s} \sup_{\rho_\mathrm{coh}(s,0,n)} \chi[\Phi|\rho_\mathrm{coh}(s,0,n)] \, .
\end{equation}

We can hence restrict our attention to transmitters of the form $\rho_\mathrm{coh}(s,0,n)$ which, 
according to Eq.~(\ref{channel}), are mapped into
\begin{equation}
\rho_{\mathrm{coh},u}(s,0,n) = \bigotimes_{k=1}^s \sigma_u(\alpha_k) \, ,
\end{equation}
with
\begin{equation}\label{thenoise}
\sigma_u(\alpha_k) = \int d\nu_k G_{n_\mathrm{th}}(\nu_k) |z_u \alpha_k + \nu_k \rangle\langle z_u \alpha_k + \nu_k| \, ,
\end{equation}
where $G_{n_\mathrm{th}}(\nu_k)$ denotes a Gaussian probability density distribution with zero mean and
variance $n_\mathrm{th}$.
Then, we notice that it is always possible to find a $u$-independent unitary matrix $U$
such that, given the energy constraint, the coherent-state amplitudes in Eq.~(\ref{thenoise})
transform as follows
\begin{equation}
\sum_{k=1}^s U_{ik} ( z_u \alpha_k + \nu_k ) = \delta_{1i} z_u \sqrt{n} + \nu'_i \, ,
\end{equation}
where the random variables $\nu'_i$'s are independent and identically distributed according
to a Gaussian distribution with zero mean and variance $n_\mathrm{th}$.
The unitary transformation $U$ on the coherent-state amplitudes can be physically implemented
by a network of passive linear-optical elements, as beam-splitters and phase-shifter.
Due to the unitary invariance of the von Neumann entropy such a transformation cannot change the
value of the Holevo information, hence the final state of the transmitter can be assumed,
without loss of generality, to be of the form
\begin{equation}
\rho_{\mathrm{coh},u}(s,0,n) = \sigma_u(\sqrt{n}) \otimes \sigma(0)^{\otimes (s-1)} \, ,
\end{equation}
where $\sigma(0) = \int d\nu G_{n_\mathrm{th}}(\nu) |\nu \rangle\langle \nu|$ is a $u$-independent
thermal state.
We notice that such a state is the tensor product of a state of the first
probing mode and a $u$-independent state of the remaining $(s-1)$ modes.
Once again, the subadditivity Holevo information implies that the presence of the 
$(s-1)$ probing modes cannot increase the Holevo function.
In conclusion, we have obtained that a single-mode coherent state
is optimal among coherent state transmitters, that is,
\begin{equation}
C_\mathrm{coh}(\Phi|n) = \chi[\Phi|\rho_\mathrm{coh}(1,0,n)] \, ,
\end{equation}
where we can assume without loss of generality $\rho_\mathrm{coh}(1,0,n) = |\sqrt{n}\rangle\langle\sqrt{n}|$.

We now consider the case of generic classical transmitters.
First of all, since coherent states are classical states,
\begin{equation}
C_\mathrm{c}(\Phi|n) \geq C_\mathrm{coh}(\Phi|n) \, .
\end{equation}
On the other hand, any classical state $\rho_c(s,r,n)$ can be 
written as the convex sum of coherent states,
\begin{equation}
\rho_\mathrm{c}(s,r,n) \hspace{-0.1cm} = \hspace{-0.15cm} \int \hspace{-0.1cm} dy p_y \bigotimes_{k=1}^s |\alpha_k(y)\rangle\langle\alpha_k(y)| \bigotimes_{{k'}=1}^r |\beta_{k'}(y)\rangle\langle\beta_{k'}(y)| \, ,
\end{equation}
with $p_y \geq 0$ and $\int dyp_y = 1$. Using the convexity of the Holevo information we get
\begin{equation}\label{convex-c-coh}
C_\mathrm{c}(\Phi|n) \leq \int dy p_y C_\mathrm{coh}(\Phi|n_y) \, ,
\end{equation}
where $n_y = \sum_{k=1}^s |\alpha_k(y)|^2$ with $n = \int dy p_y n_y$.
It is worth noticing that Eq.~(\ref{convex-c-coh}) does not suffice to
conclude that $C_\mathrm{c}(\Phi|n) \leq C_\mathrm{coh}(\Phi|n)$.
As already discussed in~\cite{QRC}, a sufficient condition for having
$C_c(\Phi|n) \leq C_\mathrm{coh}(\Phi|n)$ is that
$C_\mathrm{coh}(\Phi|n)$ is in turn a concave function of $n$.

In the remainder of this section, we focus on the case of binary encoding where
the marginal cell $\Phi \equiv \{ p_0,p_1 , z_0,z_1 \}$ is identified by the
probability weights $p_0$, $p_1=1-p_0$, and by the complex valued attenuation factors $z_0$, $z_1$. 
In the noiseless case, $n_\mathrm{th}=0$, an analytical expression for
$C_\mathrm{coh}(\Phi|n)$ has been computed in~\cite{QRC} for (real) positive values of $z_0$, $z_1$.
The latter is immediately extended to generic values of the attenuation factors, yielding
\begin{equation}\label{coh_nl}
C_\mathrm{coh}(\Phi|n) = h_2 \left[ \frac{1}{2} - \frac{1}{2}\sqrt{1-4p_0p_1(1-e^{-n |\Delta z|^2})} \right] \, ,
\end{equation}
where $\Delta z = z_1-z_0$ and 
\begin{eqnarray}
h_d[q_1, \dots, q_{d-1}] & = & - \left(1-\sum_{i=1}^{d-1} q_i\right) \log_2{\left(1-\sum_{i=1}^{d-1} q_i\right)} \nonumber \\
& - & \sum_{i=1}^{d-1} q_i \log_2{q_i}
\end{eqnarray}
is the $d$-ary Shannon entropy.
The expression in~(\ref{coh_nl}) is a concave function of $n$, hence implying that coherent
states are optimal among classical transmitters in the noiseless
case, i.e., $C_c(\Phi|n) = C_\mathrm{coh}(\Phi|n)$.
For the noisy case, $n_\mathrm{th} > 0$, we are not able to
provide an analytical expression, however $C_\mathrm{coh}(\Phi|n)$
can be easily computed numerically. The numerical evaluation of
$C_\mathrm{coh}(\Phi|n)$ suggests that it is indeed a concave
function of $n$. This leads us to conjecture that
$C_\mathrm{c}(\Phi|n) = C_\mathrm{coh}(\Phi|n)$ even in the noisy
case, i.e., coherent states are optimal among classical transmitters
even for $n_\mathrm{th} > 0$.

An approximate analytical expression for $C_\mathrm{coh}(\Phi|n)$ can be obtained in
the limit of faint signals, $n \ll 1$, where we can approximate
$|\sqrt{n}\rangle \simeq \sqrt{1-n}|0\rangle + \sqrt{n}|1\rangle$. 
Assuming $n_\mathrm{th} \ll 1$ (which is realistic in standard setups of optical
reading~\cite{Pirs}) we get to the lowest orders in $n_\mathrm{th}$,
\begin{equation}\label{Ccoh}
C_\mathrm{coh}(\Phi|n) \simeq 
h_2\left[p_0p_1 n |\Delta z|^2+n_\mathrm{th}\right]
- h_2\left[n_\mathrm{th}\right] \, .
\end{equation}
This approximate expression for $C_\mathrm{coh}(\Phi|n)$ is a concave function of $n$, 
hence we conclude that, within this approximation, $C_\mathrm{c}(\Phi|n) = C_\mathrm{coh}(\Phi|n)$.
Furthermore, retaining only the leading terms in $n \log_2{n}$ and $n_\mathrm{th} \log_2{n_\mathrm{th}}$, Eq.~(\ref{Ccoh}) yields
\begin{eqnarray}
C_\mathrm{coh}(\Phi|n) & \hspace{-1mm} \simeq & \hspace{-1mm}
n_\mathrm{th} \log_2{n_\mathrm{th}} \nonumber \\
& \hspace{-1mm} - & \hspace{-1mm} \left( p_0p_1 n |\Delta z|^2 + n_\mathrm{th} \right) \log_2{\hspace{-1mm}\left( p_0p_1 n |\Delta z|^2 + n_\mathrm{th} \right)} \, . \nonumber \\
\label{Ccoh_1}
\end{eqnarray}

Going beyond the set of classical transmitters we consider the state 
$\rho_\mathrm{EPR}(s,s,n) = (|\xi\rangle\langle\xi|)^{\otimes s}$ as an exemplary
non-classical transmitter. 
In the noiseless limit, $n_\mathrm{th}=0$, an analytical expression can be
easily derived for $|z_0| = |z_1| = 1$. 
In that case information is encoded in the relative phase $e^{i\theta}$ of the 
reflected modes, yielding
\begin{equation}\label{ent_nl}
\chi[\Phi|(|\xi\rangle\langle\xi|)^{\otimes s}] = h_2 \left[ \frac{1}{2} - \frac{1}{2}\sqrt{1-4p_0p_1\left(1-q_{n,s,\theta}\right)} \right] \, ,
\end{equation}
where
\begin{equation}
q_{n,s,\theta} = \left| 1 + \frac{n}{s}\left( 1 - e^{i\theta} \right) \right|^{-2s} \, .
\end{equation}
A comparison with~(\ref{coh_nl}) shows that a `quantum advantage' in the readout
rate, i.e., $\chi[\Phi|(|\xi\rangle\langle\xi|)^{\otimes s}] - C_\mathrm{coh}(\Phi|n) > 0$,
can be always attained (by choosing a sufficiently big $s$) when $\theta < \pi/2$, otherwise 
coherent states are always optimal probes.
The situation changes if one considers generic values of $z_0$, $z_1$~\cite{Pirs,QRC}
and if background thermal noise is added. 
Under these more general conditions one can obtain an approximate expression in the limit $n \ll 1$
by truncating the EPR state as
$|\xi\rangle \simeq \sqrt{1-n/s}|0\rangle + \sqrt{n/s}|1\rangle_{S_k} |1\rangle_{R_k}$ and
$|\xi\rangle^{\otimes s} \simeq \sqrt{1-n}|0\rangle + \sqrt{n}|\aleph\rangle$ where $|\aleph\rangle = s^{-1/2} \sum_{k=1}^s |1\rangle_{S_k} |1\rangle_{R_k}$. 
If also $n_\mathrm{th} \ll 1$, a straightforward calculation leads to the
following expression for the maximum readout rate, at the lowest orders in $n_\mathrm{th}$,
\begin{eqnarray}\label{chitb}
\chi[\Phi|(|\xi\rangle\langle\xi|)^{\otimes s}] & \simeq &
h_4[ p_0p_1 n |\Delta z|^2 , \left(1-\langle |z^2| \rangle\right) n , n_\mathrm{th}] \nonumber \\
& - & \sum_{u=0}^1 p_u h_3[ \left(1-|z_u|^2\right) n , n_\mathrm{th}] \, ,
\end{eqnarray}
where $\langle |z|^2 \rangle = p_0 |z_0|^2 + p_1 |z_1|^2$.
By retaining only the leading terms in $n \log_2{n}$ and $n_\mathrm{th} \log_2{n_\mathrm{th}}$
we finally obtain
\begin{equation}\label{chitb_1}
\chi[\Phi|(|\xi\rangle\langle\xi|)^{\otimes s}] \simeq - p_0p_1 n \left| \Delta z \right|^2 \log_2{\left( p_0p_1 n |\Delta z|^2 \right)} \, .
\end{equation}

The expressions~(\ref{chitb}),~(\ref{chitb_1}) have several remarkable properties. 
First of all, they are independent of $s$, that is, within this approximation 
the number of EPR states cannot affect the reading rate, which is
only a function of the overall mean photon number $n$. 
That implies $\chi[\Phi|(|\xi\rangle\langle\xi|)^{\otimes
s}]\simeq\chi[\Phi||\xi\rangle\langle\xi|]$ up to higher order terms. 
Notice that, as suggested by the results of~\cite{Pirs} and discussed in~\cite{QRC}, this
property does not generally hold true if $n \not\ll 1$.
The second thing to be noticed is that, contrarily to~(\ref{Ccoh_1}),
Eq.~(\ref{chitb_1}) is independent of $n_\mathrm{th}$, which implies
that the reading rate with the EPR transmitter is insensitive to
thermal-like background noise to the leading order in $n$ and $n_\mathrm{th}$.
In particular we have $\chi[\Phi||\xi\rangle\langle\xi|]/C_\mathrm{coh}[\Phi|n]\simeq \log_2{n}/\log_2{n_\mathrm{th}} > 1$ for $n < n_\mathrm{th} \ll 1$,
that is, classical light is never optimal in this regime.
On the other hand, $\chi[\Phi||\xi\rangle\langle\xi|] \simeq C_\mathrm{coh}[\Phi|n]$ for $1 \gg n \gg n_\mathrm{th}$
(implying that the quantum advantage $\chi[\Phi||\xi\rangle\langle\xi|] - C_\mathrm{coh}[\Phi|n]$ is
of higher order in this region of the parameters $n$ and $n_\mathrm{th}$).
Finally, it is worth remarking that both Eqs.~(\ref{Ccoh_1}) and~(\ref{chitb_1}) are functions of 
$p_0p_1 n \left| \Delta z \right|^2$, suggesting that the approximate
expressions hold true even if $n \not\ll 1$ as long as 
$p_0p_1 n |\Delta z|^2 \ll 1$.

\begin{figure}[tbh]
\centering
\includegraphics[width=0.48\textwidth]{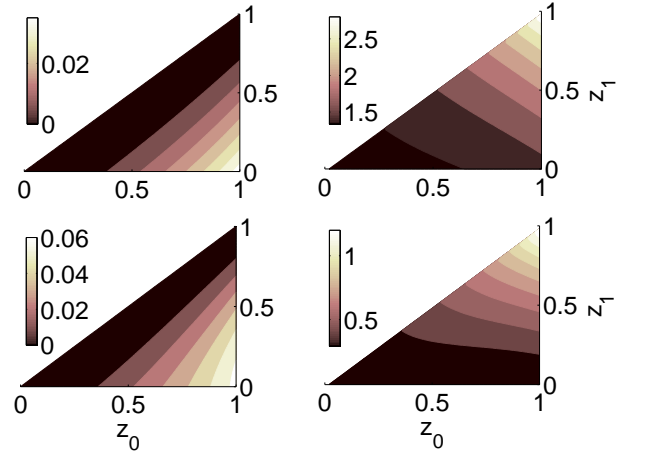}
\caption{(Color online) Density plots of the absolute information gain
$G_a$ (left)
and of the relative information gain
$G_r$ (right)
comparing the EPR transmitter with the coherent state transmitter,
as a function of $z_0$, $z_1$, for positive values of the attenuating factors ($0 \leq z_1 \leq z_0 \leq 1$). 
The top plots are for $n=0.1$, $n_\mathrm{th}=1$; 
those on the bottom are for $n=1$, $n_\mathrm{th}=1$.}
\label{1X2}
\end{figure}

For higher values of $n$, $n_\mathrm{th}$ we resort to numerical
calculations. For the sake of presentation, we put $p_0=p_1=1/2$ (this choice
of the probability maximizes both the noiseless coherent-state reading capacity ---
see Eq.~(\ref{coh_nl}) --- and the noisy one due to symmetry reasons).
Figure~\ref{1X2} shows numerical calculations of the absolute information 
gain in the reading rate,
\begin{equation}
G_a = \chi[\Phi|\rho_\mathrm{EPR}(1,1,n)]-\chi[\Phi|\rho_\mathrm{coh}(1,0,n)] \, ,
\end{equation}
and of the relative one
\begin{equation}
G_r = G_a/\chi[\Phi|\rho_\mathrm{coh}(1,0,n)] \, ,
\end{equation}
provided by the EPR transmitter over the classical transmitters, for
real-positive values of $z_0$, $z_1$.
These plots can be compared to the analogous ones in~\cite{QRC}
concerning the noiseless limit $n_\mathrm{th}=0$. 
It can be noticed that, contrarily to the noiseless case, in the noisy
setting the information gain is positive almost everywhere.
That shows, in accordance with Eqs.~(\ref{Ccoh_1}) and~(\ref{chitb_1}),
the robustness of quantum reading with the EPR transmitter
against thermal background noise.

\begin{figure}[tbh]
\centering
\includegraphics[width=0.48\textwidth]{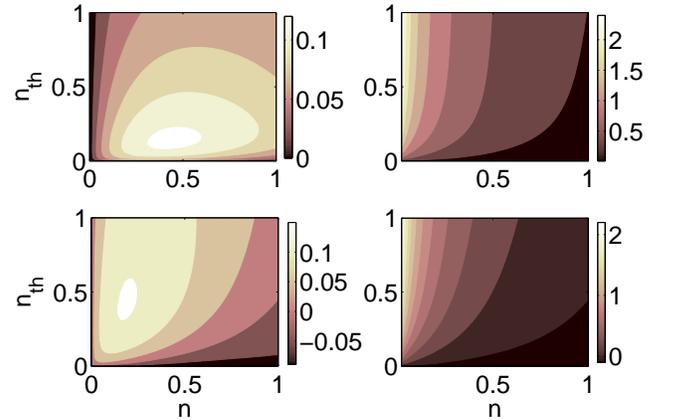}
\caption{(Color online) Density plot of the information gain
$G_a$ (left)
and of the relative information gain 
$G_r$ (right)
as a function of $n$ and $n_\mathrm{th}$
for amplitude encoding $z_0=0$, $z_1=1$ (top) 
and phase encoding $z_0=1$, $z_1=-1$ (bottom).}
\label{abs-rel}
\end{figure}

Figure~\ref{abs-rel} shows the information gain as a function of 
$n$ and $n_\mathrm{th}$ for examples of amplitude encoding
($z_0 = 0$, $z_1 = 1$) and phase encoding ($z_0=1$, $z_1=-1$).
For amplitude encoding the gain is always positive.
For phase encoding the gain is negative in the noiseless setting, that is, classical transmitter
are optimal for $n_\mathrm{th}=0$. However, the EPR transmitter always
gives better performances for sufficiently high background thermal noise.
It is worth noticing that in both amplitude and phase encoding the
relative information gain allowed by the EPR transmitter is maximum for 
$n \ll n_\mathrm{th}$ in accordance with the expressions in~(\ref{Ccoh_1}) and~(\ref{chitb_1}).

\section{Diffraction-induced interbit interference}\label{sec:interbit}

In order to retrieve information, the probe and ancillary modes
have to be collected and jointly measured. Any measurement setup
must include suitable optical components to focus the incoming
light on the surface of the detector. Such an optical system is
characterized by its finite numerical aperture~\cite{Fourier}
which induces losses and diffraction. Because of the interference
of the beams of light scattered by neighboring memory cells, the
corresponding signals overlap on the detector
surface~\cite{diffraction}, leading to a kind of correlated error
in the reading process known as {\it inter-bit
interference}~\cite{BDM,RMP}. Our aim is hence to study to what extent
this kind of error limits the efficacy of the readout protocols.
In order to pursue a first quantitative analysis of this
kind of noise we evaluate bounds on the reading capacity in the 
presence of diffraction.

We model the optical system as a thin converging lens, with
radius $R$ and focal distance $f$.
Under focusing conditions, the diffraction pattern produced
by the optical system is characterized by the associated
Rayleigh length, $x_\mathsf{R} := \lambda D_o/R$, where $\lambda$
is the wavelength of a monochromatic probing light, and $D_o$ is the
distance between the memory cells and the lens.
In this setting, the optical system is characterized by a set of
input `normal modes', associated with the canonical operators
$\{ A_i, A_i^\dag\}_{i=-\infty,\dots,\infty}$ which are independently
transmitted and attenuated by the factors $\{ t_i \}_{i=-\infty,\dots,\infty}$
(see Appendix~\ref{diffract}).
Thus, the optical system maps the input `normal modes' into a corresponding
set of output modes
$\{ B_i, B_i^\dag \}_{i=-\infty,\dots,\infty}$ according to the relations
\begin{equation}\label{independent}
B_i = t_i \, A_i + \sqrt{1-|t_i|^2} \, E_i \, ,
\end{equation}
where $\{ E_i, E_i^\dag \}_{i=-\infty, \dots, \infty}$ are vacuum modes.

A monochromatic mode $\{ a_j, a_j^\dag \}$ probing the $j$th memory cell
is scattered into a reflected mode, denoted by $\{ a^\prime_j, {a^\prime_j}^\dag \}$,
with
\begin{equation}
a'_j = z_{u(j)} \, a_j + \sqrt{1-|z_{u(j)}|^2} \, v_j + \nu_j \, .
\end{equation}
Moreover, we assume that the memory cells are located along a
straight line of length $L$ on the surface of the optical memory.
The $j$th memory cell is at positions $j \ell$, where $\ell$ denotes the spacing between
neighboring cells.
In general, the scattered modes do not form a complete set.
However, they can be completed by defining a suitable set of (possibly infinite)
normal modes $\{ e_k , e_k^\dag \}$ (assumed to be in the vacuum state).
The input normal modes $A_i$'s can hence be expanded as
\begin{equation}\label{matrix}
A_i = \sum_{j} \mathcal{M}_{ij} \, a'_j + \sum_{k} \mathcal{N}_{ik} \, e_k \, ,
\end{equation}
for suitable matrices of coefficients $\mathcal{M}_{ij}$ and $\mathcal{N}_{ik}$.
It follows that
\begin{equation}
B_i = t_i \sum_{j} \mathcal{M}_{ij} \, a'_j + t_i \sum_{k} \mathcal{N}_{ik} \, e_k + \sqrt{1-|t_i|^2} \, E_i \, .
\end{equation}
In the far-field and near-field regimes it is possible to estimate
the values of the attenuating factors~\cite{diffraction}. Here we
consider the near-field regime, in which $t_i \simeq 1$ for $|i| <
L/x_\mathsf{R}$, and $t_i \simeq 0$ for $|i| > L/x_\mathsf{R}$
(see Appendix~\ref{diffract}), yielding, for $|i| <
L/x_\mathsf{R}$,
\begin{equation}
B_i = \sum_{j} \mathcal{M}_{ij} \, a'_j + \sum_{k} \mathcal{N}_{ik} \, e_k \, .
\end{equation}

Since the matrix $\mathcal{M}$ has in general non-zero off-diagonal terms, the light beams
reflected by different memory cells do overlap at the surface of
the detector.
To analyze this phenomenon, we introduce the singular value decomposition,
\begin{equation}
\mathcal{M}_{ij} = \sum_{h} \, \mathcal{V}^*_{hi} \, \tau_h \, \mathcal{U}_{hj} \quad (\mbox{for} \, \,  |i| < L/x_\mathsf{R}) \, ,
\end{equation}
where $\mathcal{U}$ and $\mathcal{V}$ are unitary matrices, and $\tau_h \geq 0$ are the (squared)
singular values of $\mathcal{M}$. We can hence introduce the new sets of canonical
modes $\tilde{B}_h:=\sum_{|i| < L/x_\mathsf{R}} \mathcal{V}_{hi} B_i$, and
$\tilde{a}^\prime_h:=\sum_j \mathcal{U}_{hj} a^\prime_j$, in terms of which the
transformation becomes diagonal,
\begin{equation}\label{unraveled}
\tilde{B}_h = \tau_h \, \tilde{a}^\prime_h + \sum_{|i| < L/x_\mathsf{R},k} \, \mathcal{V}_{hi} \, \mathcal{N}_{ik} e_k \, .
\end{equation}
Finally, we notice that consistency with the canonical commutation relations
for the operators $\{ \tilde{B}_h, \tilde{B}_h^\dag \}$ enforces that
$\sum_{|i| < L/x_\mathsf{R},k} \mathcal{V}_{hi} \mathcal{N}_{ik} e_k = \sqrt{1-|\tau_h|^2} \, \tilde{e}_h$,
where $\{ \tilde{e}_h, \tilde{e}_h^\dag \}$ is a suitably defined set of vacuum modes.
Thus, Eq.~(\ref{unraveled}) expresses the fact that the modes
$\tilde{a}^\prime_h$ are independently attenuated, with attenuating
factors $\{ \tau_h \}$.

It is worth remarking that the independently attenuated modes
$\tilde{a}^\prime$'s do not coincide with the modes $a^\prime$'s carrying the
information on the state of the individual memory cells.
This is the very effect of diffraction, which mixes the signal scattered
by different memory cells. As a consequence, the state $\rho_{\mathbf{u}^i}(s,r)$ in
Eq.~(\ref{ostate}) has to be replaced with a state of the form
$\rho'_{\mathbf{u}^i}(s,r) = ( \mathcal{E}_{Ks} \otimes I^{\otimes Kr} ) \rho_{\mathbf{u}^i}(s,r)$,
where $\mathcal{E}_{Ks}$ is the map describing the attenuation of the signal modes
$\tilde{a}^\prime$'s, and $I^{\otimes Kr}$ is the identity map on the remaining
$Kr$ ancillary modes. For the sake of presentation, here we assume that
diffraction only involves the signal modes. The case of diffraction on both
the signal and ancillary modes is simply obtained by substituting the map
$( \mathcal{E}_{Ks} \otimes I^{\otimes Kr} )$ with $( \mathcal{E}_{Ks} \otimes \mathcal{E}_{Kr} )$.
In both cases, the tensor-product structure of Eq.~(\ref{ostate}) is lost.
This prevents us from expressing the reading rate in terms of a `single-letter'
expression, as in Eq.~(\ref{HI}).

Nevertheless, upper and lower bounds can be estimated by
exploiting the `data processing inequality' for the Holevo information,
which implies that the extra noise term expressed by the attenuating
channel $\mathcal{E}_{Ks}$ can only reduce the reading rates.
Such a reduction is in a range determined by the maximum and minimum values
of the attenuating factors $\{ \tau_h \}$, denoted by $\tau_\mathrm{max}$
and $\tau_\mathrm{min}$.
We hence consider the fictitious channels $\mathcal{E}_\mathrm{max}$ and $\mathcal{E}_\mathrm{min}$
which independently attenuate all the modes $\tilde{a}^\prime$'s by factors $\tau_\mathrm{max}$
and $\tau_\mathrm{min}$ respectively.
As a matter of fact, due to the linear relation between the modes $\tilde{a}^\prime$'s
and the modes $a^\prime$'s, the latter are also independently attenuated by the maps
$\mathcal{E}_\mathrm{max}$ and $\mathcal{E}_\mathrm{min}$.
It follows that the fictitious channels do preserve the tensor product structure of
Eq.~(\ref{ostate}), hence allowing us to employ the single-letter formula for the
reading rate.
We can hence write the following bounds for the reading rate $R[\rho(s,r)]$ in the presence of diffraction:
\begin{equation}\label{ulbounds}
\chi[\Phi_\mathrm{min}|\rho(s,r,n)] \leq R[\rho(s,r,n)] \leq \chi[\Phi_\mathrm{max}|\rho(s,r,n)] \, ,
\end{equation}
where $\Phi_\mathrm{min} = \{ p_u , \phi_{\mathrm{min},_u} \}$ and
$\Phi_\mathrm{max} = \{ p_u , \phi_{\mathrm{max},_u} \}$ are the
channel ensembles obtained by composition of the ensemble 
$\Phi = \{ p_u , \phi_u \}$ with the single-mode attenuating channels with
attenuating factors $\tau_\mathrm{min}$ and
$\tau_\mathrm{max}$, respectively.
Explicitly, the channels $\phi_{\mathrm{min},_u}$,
$\phi_{\mathrm{max},_u}$ respectively transform a probing mode $\{ a, a^\dag \}$ 
according to
\begin{equation}
a \to \tau_\mathrm{min} z_{u(j)} \, a + \sqrt{1-|\tau_\mathrm{min}z_{u(j)}|^2} \, v + \tau_\mathrm{min} \nu \, ,
\end{equation}
and
\begin{equation}
a \to \tau_\mathrm{max} z_{u(j)} \, a + \sqrt{1-|\tau_\mathrm{max}z_{u(j)}|^2} \, v + \tau_\mathrm{max} \nu \, .
\end{equation}

\subsection{Bounds on the reading rates}\label{taus}

In this section, we estimate the upper and lower bounds in~(\ref{ulbounds}) on the reading rates
in the presence of diffraction by computing the maximum and minimum
of the attenuation factors $\tau_h$'s.
We model the optical system associated with the measurement device
as a converging thin lens of radius $R$, and assume that the memory
cells are located along a straight line on the surface of the optical memory,
where each cell has linear extension $d$, and neighboring cells are spaced by 
$\ell$ for a total length equal to $L$.
A characterization of the light propagation inside such an optical system is
presented in Appendix~\ref{diffract} (based on~\cite{diffraction}).

We assume that the mode describing the light reflected at the $j$th memory cell, on
the surface of the optical memory, is of the form
\begin{equation}
a^\prime_j = \frac{1}{\sqrt{d}} \int_{j\ell-d/2}^{j\ell+d/2} dx A(x) \, , \\
\end{equation}
where $x$ is a linear coordinate at the memory surface, and the
continuous set of operators $\{ A(x), A^\dag(x) \}$ corresponds to the
quantized amplitudes of the electromagnetic field on the surface.
The latter can be in turn expressed in terms of the discrete set $\{ A_i, A_i^\dag\}$ of
Fourier-transformed modes on the line [defined in Eq.~(\ref{AA})].
These are the input `normal modes' which are independently attenuated by the optical system.
We hence obtain
\begin{align}
a^\prime_j & = \sum_{i=-\infty}^\infty A_i \int_{j\ell-d/2}^{j\ell+d/2} \frac{dx}{\sqrt{d L}} \exp{\left[ \mathrm{i} 2\pi i \frac{x}{L} + \mathrm{i} \theta_o(x) \right]} \nonumber \\
& \simeq \sum_{i=-\infty}^\infty A_i e^{\mathrm{i} \theta_o(j\ell)} \int_{j\ell-d/2}^{j\ell+d/2} \frac{dx}{\sqrt{d L}} \exp{\left( \mathrm{i} 2\pi i \frac{x}{L} \right)} \nonumber \\
& = \sqrt{\frac{d}{L}} \, e^{\mathrm{i}\theta_o(j\ell)} \sum_{i=-\infty}^\infty
\frac{\sin{(\pi i d/L)}}{\pi i d/L} \exp{\left( \mathrm{i} 2\pi j i \frac{\ell}{L} \right)} A_i \, ,
\end{align}
where $\mathrm{i}$ stands for the imaginary unit and $\theta_o(x)$ is a position-dependent 
phase factor, which can be assumed to be constant on the intervals
$[j\ell-d/2,j\ell+d/2]$ as long as $d^2 \ll \lambda D_o$ (see Appendix~\ref{diffract}).
From that we get the expression for the (adjoint of
the) matrix of Eq.~(\ref{matrix})
\begin{equation}
\mathcal{M}_{ij}^* = \sqrt{\frac{d}{L}} \, e^{\mathrm{i}\theta_o(j\ell)}
\frac{\sin{(\pi i d/L)}}{\pi i d/L} \exp{\left( \mathrm{i} 2\pi j i \frac{\ell}{L} \right)} \, .
\end{equation}
This is a rectangular semi-infinite matrix [with $j=-L/(2\ell), \dots, L/(2\ell)$]
that we will truncate by restricting it to the modes $A_i$'s actually transmitted
across the optical systems. In the near-field regime this corresponds to restrict the range
to $i=-L/x_\mathsf{R}, \dots, L/x_\mathsf{R}$, where
$x_\mathsf{R}:=\lambda D_o/R$ is the Rayleigh number (see Appendix~\ref{diffract}).

The (squared) singular values of the matrix $\mathcal{M}$ are
the eigenvalues of the Hermitian matrix $\mathcal{M} \mathcal{M}^\dag$, with elements:
\begin{equation}
(\mathcal{M} \mathcal{M}^\dag)_{kj} = \frac{d}{L}
\sum_i \left[\frac{\sin{(\pi i d/L)}}{\pi i d/L}\right]^2
\exp{\left[ \mathrm{i} 2\pi (j-k) n \frac{\ell}{L}
\right]} \, .
\end{equation}
In the limit $L/x_\mathsf{R} \gg 1$ and $\ell/L \ll 1$, we get
\begin{equation}
(\mathcal{M} \mathcal{M}^\dag)_{kj} = \frac{d}{\ell}
\int_{-\ell/x_\mathsf{R}}^{\ell/x_\mathsf{R}} dx
\left[\frac{\sin{(\pi x d/\ell)}}{\pi x d/\ell}\right]^2
e^{\mathrm{i} 2\pi (j-k) x} \, .
\end{equation}
The matrix $\mathcal{M} \mathcal{M}^\dag$ is a Toeplitz matrix,
with entries only depending on the difference $j-k$. The spectrum
of a Toeplitz matrix is bounded by the maximum and the minimum of
the Fourier transform~\cite{Gray}
\begin{equation}
f(z) = \sum_q (\mathcal{M} \mathcal{M}^\dag)_q \, e^{-\mathrm{i} z q} \, ,
\end{equation}
for $z \in [0,2\pi]$, where $(\mathcal{M} \mathcal{M}^\dag)_q := (\mathcal{M} \mathcal{M}^\dag)_{kj}$ for
$j-k=q$. Notice that the integer $q$ varies in the range $q \in [-L/\ell,-L/\ell]$.
In the limit $L/\ell \gg 1$, the summation over $q$ can be extended up to $\pm \infty$, yielding
\begin{align}
f(z) & \simeq \sum_{q=-\infty}^\infty \frac{d}{\ell}
\int_{\frac{-\ell}{x_\mathsf{R}}}^{\frac{\ell}{x_\mathsf{R}}} dx
\left[\frac{\sin{(\pi x d/\ell)}}{\pi x d/\ell}\right]^2
e^{\mathrm{i} (2\pi x - z) q} \nonumber\\
& = \frac{d}{\ell} \int_{\frac{-\ell}{x_\mathsf{R}}}^{\frac{\ell}{x_\mathsf{R}}} dx
\left[\frac{\sin{(\pi x d/\ell)}}{\pi x d/\ell}\right]^2
\sum_{m=-\infty}^\infty \delta\left(x + m - \frac{z}{2\pi}\right) \nonumber\\
& = \sum_{m=y/(2\pi)-\ell/x_\mathsf{R}}^{y/(2\pi)+\ell/x_\mathsf{R}} \frac{d}{\ell}
\left[\frac{\sin{(\pi (y/(2\pi)-m) d/\ell)}}{\pi (y/(2\pi)-m) d/\ell}\right]^2 \, .
\end{align}
From the extrema of this function one can finally compute the
attenuating factors $\tau_\mathrm{min}$, $\tau_\mathrm{max}$ and the bounds
$\chi[\Phi_\mathrm{min}|\rho(s,r,n)]$, $\chi[\Phi_\mathrm{max}|\rho(s,r,n)]$.

\begin{figure}[tbh]
\centering
\includegraphics[width=0.48\textwidth]{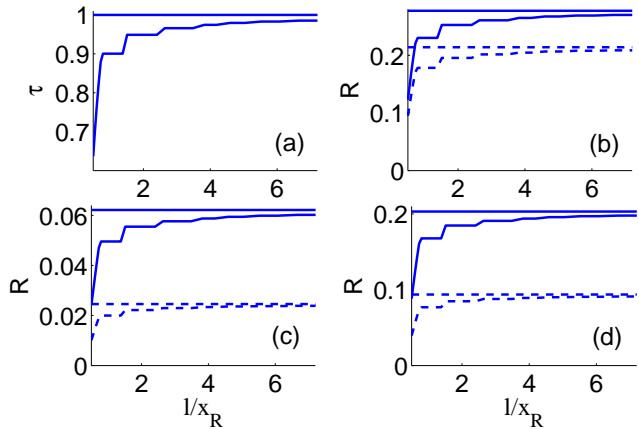}
\caption{The plot shows as a function of the ratio $\ell/x_\mathsf{R}$:
(a) the attenuating factors $\tau_\mathrm{min}$ and $\tau_\mathrm{max}$;
(b), (c), (d) the lower and upper bounds on the reading rates for $p_0=p_1=1/2$ 
with the EPR transmitter (solid lines) and the coherent-state transmitter (dashed lines).
In (b) $z_0=0$, $z=1$, $n=1$, $n_\mathrm{th}=1$; 
in (c) $z_0=0$, $z_1=1$, $n=0.1$, $n_\mathrm{th}=1$; 
in (d) $z_0=1$, $z_1=-1$, $n=0.1$, $n_\mathrm{th}=1$.} 
\label{diffraction}
\end{figure}

In particular, we consider the binary marginal cell $\Phi \equiv \{ p_0,p_1, z_0,z_1 \}$
with $p_0=p_1=1/2$ and real attenuation factors and
compute the bounds on the classical reading
rate by putting $\rho_\mathrm{coh}(1,0) =
|\sqrt{n}\rangle\langle\sqrt{n}|$ to be a single-mode coherent
state of amplitude $\sqrt{n}$. 
As an example of a nonclassical
transmitter, we also compute the bounds on the reading rate by
probing with a single EPR state,
$\rho_\mathrm{EPR}(1,1)=|\xi\rangle\langle\xi|$. 
In the region with $n, n_\mathrm{th} \ll 1$, these bounds can 
be computed according to Eqs.~(\ref{Ccoh}) and (\ref{chitb}). 
Outside of this region, we can still compute exactly the bounds 
on the classical reading capacity for $n_\mathrm{th}=0$~\cite{QRC}, 
otherwise we have to rely on numerical calculations. 
Figure~\ref{diffraction} shows the bounds, for $d/\ell=1$, as a
function of the dimensionless parameter $\ell/x_\mathsf{R}$. 
The limit $\ell/x_\mathsf{R} \gg 1$ corresponds to the diffraction-free 
setting analyzed in Sec.~\ref{sec:QThermal} in which $(\mathcal{M} \mathcal{M}^\dag)_{kj} \simeq \delta_{kj}$. 
On the other hand, in the region $\ell/x_\mathsf{R} \simeq 1$ 
the effects of diffraction become sensible. 
The choice $d/\ell=1$ is optimal as it maximizes the value of $\tau_\mathrm{max}$, 
yielding $\tau_\mathrm{max}=1$. 
Notice that in the high-diffraction region $\ell/x_\mathsf{R}<1/2$, the lower 
bound $\chi[\Phi_\mathrm{min}|\rho(s,r,n)]$ becomes trivial, since $\tau_\mathrm{min}=0$. 
The separation between the lower bound for the EPR state probing and the upper 
bound for the classical reading capacity shows that the information gain in the 
reading capacity persists in the presence of interbit interference caused by light
diffraction.

\section{Conclusions}\label{conclusions}

Quantum reading exploits quantum features of light to boost the statistical 
discrimination of bosonic channels, which is the essential mechanism in
the readout of digital optical memories. 
The advantages of quantum light over the classical one can be remarkable in the region of low
power per pulse, a feature that may lead to technological applications for a
high-density data storage device and fast memory readout. 
Moreover, it may
allow noninvasive and reliable utilization of data storage devices based on
photodegradable materials.

We have analyzed the quantum reading capacity in the presence of
thermal background noise and correlated noise arising from
diffraction of the probing light. We have shown that
probing the memory with EPR states allows a reading rate which is,
contrarily to classical states of light, mostly insensitive to
thermal noise in the regime of low power per pulse. 
This feature mirrors the quantum advantage allowed by {\it quantum
illumination} for the problem of target detection~\cite{illumination}
and conveys the physical mechanism at the root of a class of two-way 
cryptographic protocols~\cite{Shapiro,Exp}
(see Ref.~\cite{SSLB} for general protocols of two-way quantumcryptography
with continuous-variable systems).

Potential applications for high-density data storage technologies
motivated us to study quantum reading under the effects of
diffraction, which induces correlated noise in the reading process
(interbit interference). By modeling the diffraction as caused by
a linear optical system characterized by its Rayleigh number, we
have shown that the advantage of quantum probing over the classical
one persists in the presence of this kind of correlated noise.

The problem remains open of designing optimal codes and collective measurements
(allowing one to achieve the optimal rates expressed by the Holevo information) 
which can be experimentally implemented with current technologies.
This is in particular true in the presence of intersymbol interference caused 
by diffraction, in which case one has to face a strategy to counteract cross talks
among different modes.
Remarkably, an explicit capacity-achieving receiver for quantum reading has been 
put forward in the noiseless limit for the case in which binary information is 
encoded in phase, that is $|z_u|=1$~\cite{Mark}.


\acknowledgments C.L and S.P. would like to thank S. Guha for helpful comments.
The research leading to these results has received
funding from the European Commission's seventh Framework Programme
(FP7/2007-2013) under grant agreements no.~213681, by the
Italian Ministry of University and Research under the FIRB-IDEAS
project RBID08B3FM, and by EPSRC under the research grant HIPERCOM (EP/J00796X/1). 

\appendix

\section{Modeling diffraction}\label{diffract}

Here we model the effects of diffraction by assuming that the
probing light is collected and focused on the detector through a
converging linear optical system.
The latter is modeled as a thin lens of focal length $f$, which is
at distances $D_o$, $D_i$ from the object plane (the memory surface)
and the image plane (the surface of the detector) respectively.
The focusing condition is expressed by the lens law, $1/D_o + 1/D_i = 1/f$.

We assume that memory cells are disposed along a straight line on
the surface of the optical memory.
Let us consider the amplitude of the classical (scalar) electromagnetic field
at wavelength $\lambda$, $A(x_o)$, where $x_o$ is a Cartesian coordinate
on the line at the memory surface, and the amplitude $B(x_i)$, with $x_i$ a
coordinate on the corresponding line on the detector surface.
For monochromatic light, the amplitude on the object and image plane are
related by the relations~\cite{Fourier}
\begin{equation}\label{linear}
B(x_i) = \int dx_o \, T(x_i,x_o) \, A(x_o) \, ,
\end{equation}
where the point-spread function is
\begin{equation}\label{transfer}
T(x_i,x_o) = e^{\mathrm{i}\theta}
\int \frac{dx P(x)}{\lambda\sqrt{D_oD_i}} \exp{\left[ -\mathrm{i} 2\pi \frac{(x_i/M-x_o)x}{\lambda D_o}
\right]} \, ,
\end{equation}
$M=D_i/D_o$ is the magnification factor, and $\theta=\theta(x_i,x_o)=\theta_o(x_o)+\theta_i(x_i)$
with $\theta_o(x_o) = \pi |x_o|^2/(\lambda D_o) + 2 \pi D_o/\lambda$,
$\theta_i(x_i) = \pi |x_i|^2/(\lambda D_i) + 2 \pi D_i/\lambda$.
The function $P(x)$ is the characteristic function of the entrance pupil of the optical system.
We consider a slit-shaped entrance pupil, characterized by
\begin{eqnarray}
P(x) = \left\{ \begin{array}{lcr}
1 & \mbox{for} & |x|<R \, , \\
0 & \mbox{for} & |x|>R \, ,
\end{array}\right.
\end{eqnarray}
which yields the following expression for the point-spread function:
\begin{equation}
T(x_i,x_o) = \frac{e^{\mathrm{i}\theta(x_i,x_o)}R}{\lambda\sqrt{D_oD_i}}
\, \frac{\sin{\left[2\pi(x_i/M-x_o)/x_\mathsf{R}\right]}}{\pi(x_i/M-x_o)/x_\mathsf{R}}
\, ,
\end{equation}
where $x_\mathsf{R}:=\lambda D_o/R$ is the Rayleigh length.

For describing the propagation of light through the optical system,
we consider a line element of length $L$ at the memory surface and
its image of length $ML$ on the detector, and introduce the Fourier-transformed
field amplitudes
\begin{align}
A_{k} & := \int_{\frac{-L}{2}}^{\frac{L}{2}} \frac{dx_o}{\sqrt{L}} \exp{\left[ - \mathrm{i} 2\pi \frac{k x_o}{L} - \mathrm{i} \theta_o(x_o) \right]} A(x_o) \, , \label{AA} \\
B_{k} & := \int_{\frac{-ML}{2}}^{\frac{ML}{2}} \frac{dx_i}{\sqrt{ML}} \exp{\left[ - \mathrm{i} 2\pi \frac{k x_i}{ML} - \mathrm{i} \theta_i(x_i) \right]} B(x_i) \, ,
\end{align}
where $\theta_o(x_o)$, $\theta_i(x_i)$ are phase factors introduced to
compensate $\theta(x_i,x_o)$ in Eq.~(\ref{transfer}).
They satisfy the relation
\begin{equation}
B_{k} = \sum_{h} T_{kh} \, A_{h} \, ,
\end{equation}
where the {\it transfer matrix} has entries
\begin{equation}
T_{kh} = \frac{1}{\lambda\sqrt{D_oD_i}} \int dx P(x) \Delta_{kh}(x) \, ,
\end{equation}
with
\begin{align}
\Delta_{kh}(x) = & \frac{1}{\sqrt{M}L}
\int_{-L/2}^{L/2} dx_o \exp{\left[ \mathrm{i} 2\pi \left(\frac{h}{L}+\frac{x}{\lambda D_o}\right) x_o \right]} \nonumber \\
& \times\int_{-ML/2}^{ML/2} dx_i \exp{\left[ - \mathrm{i} 2\pi \left(\frac{k}{L}+\frac{x}{\lambda D_i}\right) x_i\right]} \, .
\end{align}

The expression for the transfer matrix simplifies in the near-field limit and far-field limits~\cite{diffraction}.
In particular, in the near-field limit, $L/x_\mathrm{R} \gg 1$ we have
\begin{equation}
\Delta_{kh}(x) \simeq \lambda \sqrt{D_oD_i} \, \delta_{kh} \,
\delta\left( x + \frac{\lambda D_o h}{L}\right) \, ,
\end{equation}
yielding
\begin{equation}\label{tdiag}
T_{kh} = t_{h} \, \delta_{kh} \, ,
\end{equation}
with
\begin{eqnarray}
t_{h} = \left\{ \begin{array}{lcr}
1 & \mbox{for} & |h|<L/x_\mathrm{R} \, ,\\
0 & \mbox{for} & |h|>L/x_\mathrm{R} \, .
\end{array}\right.
\end{eqnarray}

Upon quantization, the field amplitudes on the memory and the detector surface
are promoted to the canonical operators $\{ A_{h}, A^\dag_{h} \}$ and $\{ B_{h}, B^\dag_{h} \}$.
The form of Eq.~(\ref{tdiag}) hence implies that the modes $\{ A_{h}, A^\dag_{h} \}$ are
independently attenuated with attenuation factors $\{ t_{h} \}$, from which
Eq.~(\ref{independent}) is deduced.


\end{document}